\documentclass[prc,superscriptaddress,twocolumn]{revtex4-1}
\usepackage{graphicx}

\usepackage{geometry}
\begin{document}

\title{
Time-dependent Hartree-Fock plus Langevin approach 
for hot fusion reactions to synthesize the \emph{Z}\,=\,120 superheavy element
}

\author{K. Sekizawa}
\affiliation{Center for Transdisciplinary Research, Institute for Research Promotion, Niigata University, Niigata 950-2181, Japan}

\author{K. Hagino}
\affiliation{Department of Physics, Tohoku University, Sendai 980-8578,  Japan} 
\affiliation{Research Center for Electron Photon Science, Tohoku University, 1-2-1 Mikamine, Sendai 982-0826, Japan}


\begin{abstract}
We develop a novel approach to fusion reactions for syntheses of superheavy elements, which combines
the time-dependent Hartree-Fock (TDHF) method with a dynamical diffusion model based on the
Langevin equation. In this approach, the distance of the closest approach for the capture process
is estimated within the TDHF approach, which is then plugged into the dynamical diffusion model
as an initial condition. We apply this approach to hot fusion reactions leading to formation of the element
$Z=120$, that is, the $^{48}$Ca+$^{254,257}$Fm, $^{51}$V+$^{249}$Bk, and $^{54}$Cr+$^{248}$Cm
reactions. Our calculations indicate that the distances of the closest approach for these systems are
similar to each other and thus the difference in the probabilities of evaporation residue formation
among those reaction systems originates mainly from the evaporation process, which is sensitive to
the fission barrier height and the excitation energy of a compound nucleus.
\end{abstract}


\maketitle

The physics of superheavy elements is one of the most important topics in nuclear physics today
\cite{HM00,HHO13,NPA,Giuliani18,Witek18,Hagino19}. Using heavy-ion fusion reactions,
researchers have so far successfully synthesized the elements up to $Z$\,=\,118 \cite{NPA}.
Since the formation probability of superheavy elements is extremely small, it is crucial to choose
an appropriate reaction system, that is, a combination of a projectile and a target nuclei. For
this purpose, two different experimental strategies have been employed. One is the
$^{208}$Pb-based cold fusion reactions, for which the compound nucleus is formed with
relatively low excitation energies so that the survival probability of the compound nucleus
against fission is maximized. The other is the $^{48}$Ca-based hot fusion reactions,
for which the formation probability of the compound nucleus is maximized. 

It has been shown that the evaporation residue cross sections associated with the cold fusion
reactions drop rapidly, as the charge number $Z$ of the compound nucleus increases. Because
of this behavior, the cold fusion reactions have been limited only up to nihonium ($Z$\,=\,113) \cite{Morita12}.
On the other hand, the observed cross sections remain relatively large between $Z$\,=\,113
and 118 for hot fusion reactions \cite{HHO13}. It has been conjectured that this behavior
originates from the fact that the compound nuclei formed are in the proximity of the island 
of stability \cite{island1,island2} and/or an increase of dissipation at high temperatures
\cite{Loveland14}. For this reason, the hot fusion reactions are regarded as a promising
means to go beyond the known heaviest element, oganesson ($Z=118$), and synthesize
new superheavy elements.

To synthesize the new elements, $Z$\,=\,119 and 120, with hot fusion reactions utilizing the
$^{48}$Ca projectile as in the previous successful measurements, use of Es ($Z$\,=\,99) and
Fm ($Z$\,=\,100) targets is mandatory. However, due to the short half-lives of these elements,
they would not be available with sufficient amounts for fusion experiments \cite{Dullmann17}.
It is therefore inevitable to use heavier projectile nuclei, such as $^{50}$Ti, $^{51}$V, and $^{54}$Cr,
instead of $^{48}$Ca. An important question arises: how much are evaporation residue
cross sections altered if those heavier projectiles are used instead of the $^{48}$Ca nucleus?
In particular, one may ask how the double magic nature of $^{48}$Ca influences the evaporation
residue cross sections.

The role of magicity in fusion reactions has been demonstrated in Ref.~\cite{Satou06} for the 
$^{86}$Kr+$^{138}$Ba and $^{86}$Kr+$^{134}$Ba systems. In this experiment, it was
shown that the cross sections for the former system are systematically larger than those for the
latter. An interpretation of this behavior is that the projectile nucleus can come closer to the target
nucleus with less friction in the former system, in which the target nucleus has the $N=82$ magic
number. See also Ref.~\cite{Moller97} for single-particle energies for the $^{70}$Zn+$^{208}$Pb
system as a function of the internucleus distance. We also mention that a recent experiment for the
$^{50,52,54}$Cr+$^{204,206,208}$Pb systems \cite{Mohanto18} clearly showed that fusion
cross sections can be enhanced by the entrance-channel magicity provided that the $N/Z$
asymmetry is small, as discussed in Ref.~\cite{SimenelPLB12}. It may be natural to expect
that the $^{48}$Ca nucleus maintains a similar effect as well.

To address this question, one would need a microscopic approach based on the nucleonic
degrees of freedom with minimal assumptions on dynamics. The aim of this paper is to develop a
new hybrid model based on such a microscopic approach, for which we employ the time-dependent
Hartree-Fock (TDHF) theory. The TDHF approach is free from empirical parameters once the energy
density functional is fixed from nuclear structure calculations. In recent years, the TDHF approach
has been extensively applied to heavy-ion reactions around the Coulomb barrier (see, e.g.,
Refs.~\cite{Sekizawa19,Stevenson19,Simenel18,Nakatsukasa16,Simenel12}, for recent reviews).
Of course, the TDHF approach is valid only for the main process in a reaction \cite{Negele82,Simenel12},
and we cannot expect that the TDHF approach is able to describe the entire formation process of evaporation
residues of superheavy elements. We mention that one of the main processes in fusion for syntheses of 
superheavy elements is quasifission, which is a re-separation of  the two colliding nuclei before the compound
nucleus formation. It has been shown that the TDHF approach is applicable for its description, at least
for its average behavior \cite{Wakhle14,Umar15,Oberacker14,Sekizawa16}. The main idea of the hybrid
approach developed in this paper is to extract the distance of the closest approach from such calculations,
and use it as an initial condition of the diffusion process over the inner barrier, for which the Langevin approach
has been developed \cite{Abe02,fbd03,Aritomo04,ZG15}. We shall apply this new approach to fusion
reactions to form the element $Z$\,=\,120, and clarify the role of the magicity of the $^{48}$Ca
projectile in fusion reactions for superheavy elements. 

Conceptually, the formation process of evaporation residues can be divided into the following
three sub-processes \cite{Abe02,fbd03,Aritomo04,ZG15}. The first is the capture process, 
in which the projectile and target nuclei come close to the touching configuration. The second
is the diffusion process, in which the touching configuration undergoes the shape evolution
towards the compound nucleus by overcoming an inner barrier. At this stage, there is a strong
competition between this diffusion process and the reseparation, i.e., quasifission. The third
process is the statistical decay of the compound nucleus, in which there is a severe competition
between evaporations and fission. Cross sections for the formation of evaporation residues are
then given as a product of the probability of each of these three processes, that is,
\begin{equation}
\sigma_{\rm ER} (E)=\frac{\pi}{k^2}\sum_l(2l+1)T_l(E)P_{\rm CN}(E,l)W_{\rm sur}(E^*,l),
\label{er}
\end{equation}
where $l$ is a partial wave, $E$ is the incident energy in the center of mass frame,  and $k$
is given by $k=\sqrt{2\mu E/\hbar^2}$ with $\mu$ being the reduced mass in the entrance channel.
$T_l, P_{\rm CN}$, and $W_{\rm sur}$ are the probabilities for the first, the second, and the third
processes, respectively. The survival probability, $W_{\rm sur}$, is a function of the excitation 
energy, $E^*$, and the angular momentum, $l$, of the compound nucleus. In the following, to
compare the several systems, we focus only on the $s$-wave scattering (i.e., $l$\,=\,0)
and take the energy $E$ to be above the Coulomb barrier so that the capture probability, $T_l$, can
be set to unity.

We first compare the $^{54}$Cr+$^{248}$Cm and $^{48}$Ca+$^{254}$Fm systems,
both of which lead to the same compound nucleus, $^{302}$120, even though the $^{254}$Fm
nucleus is short lived and it cannot be used as the target nucleus in an actual experiment. The first step
in our approach is to perform the TDHF calculation for a head-on collision, $b=0$. For TDHF calculations,
we use the three-dimensional TDHF code developed by Sekizawa and Yabana (see, e.g., Refs.~\cite{
Sekizawa16,Sekizawa13,Sekizawa14,Sekizawa17} for details of the numerical implementation).
For the energy density functional, we use the Skyrme SLy4d parameter set \cite{SLy4d}. The pairing 
correlations are disregarded in this work. To obtain a spherical ground state, the filling approximation
is employed for $^{54}$Cr, filling proton $f_{7/2}$ and neutron $p_{3/2}$ orbitals with equal weights
for the magnetic substates. The $^{248}$Cm and $^{254}$Fm nuclei are prolately deformed in their
ground state. Since a dominant contribution to evaporation residue cross sections comes from the side
collision \cite{Hinde18,Hinde95,Hinde96,Nishio08,Nishio00,Tanaka18,Hagino18}, for which the projectile
nucleus collides with the equatorial side of the target, for simplicity we restrict our calculations only to
this configuration in the present paper.

\begin{figure} [tb]
\includegraphics[width=8.6cm]{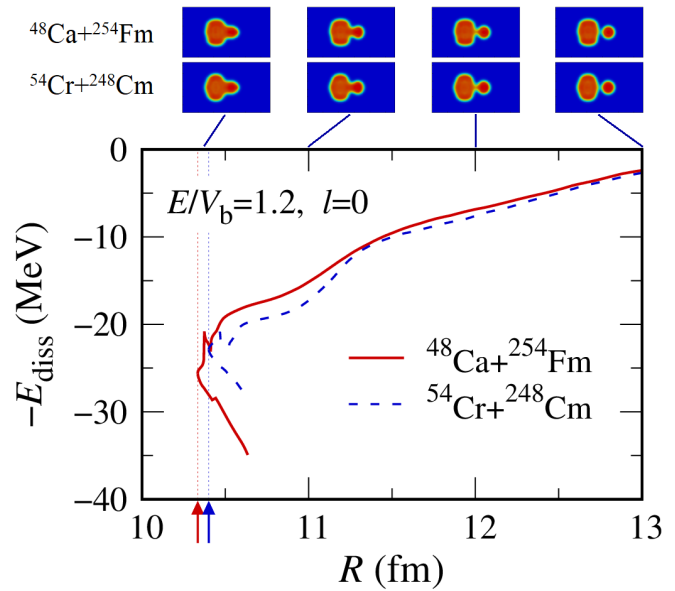}
\caption{
Results of the TDHF calculations (with the SLy4d functional \cite{SLy4d}) for the energy loss
for the relative motion between the projectile and target nuclei, shown as a function of
the internucleus distance $R$. The solid and dashed lines are for the $^{48}$Ca+$^{254}$Fm
and $^{54}$Cr+$^{248}$Cm reactions, respectively, for the $s$-wave scattering ($b=0$)
with the side collision geometry. The incident energy is set to be 1.2 times the Coulomb barrier
height for the side collision, which is evaluated with the frozen Hartree-Fock approximation.
The arrows and vertical dotted lines indicate the distance of the closest approach for each
system. At the top of the figure, the density distribution in the reaction plane is exhibited for
$^{48}$Ca+$^{254}$Fm (the top row) and $^{54}$Cr+$^{248}$Cm (the bottom row).
Each panel corresponds to a different instance at which the relative distance is $R=R_{\rm min}$,
11, 12, and 13\,fm, respectively, from left to right.
}
\label{fig1}
\end{figure}

Following Ref.~\cite{Washiyama08}, we extract from the TDHF time evolution the relative
distance, $R(t)$, between the two fragments, its conjugate momentum, $P(t)$, the reduced
mass, $\mu(R(t))$, the internucleus potential, $V(R(t))$, and the friction coefficient,
$\gamma(R(t))$, as a function of time, $t$. Having these quantities at hand, we can
compute the energy for the relative motion, 
\begin{equation}
E_{\rm rel}(R(t))=\frac{P(t)^2}{2\mu(R(t))}+V(R(t)).
\end{equation}
Notice that, because of internal excitations, the energy is dissipated from the relative
motion to internal degrees of freedom. Figure~\ref{fig1} shows the minus of the dissipative
energy, $E_{\rm diss}$, which is nothing but the relative energy with respect to the initial 
energy, $E$, for these systems as a function of the relative distance, $R$. The initial energy
is set to 1.2 times the Coulomb barrier height for the side collision. Here, the barrier height,
$V_b$, is estimated using the frozen Hartree-Fock method \cite{Simenel17}. This yields
$V_b$\,=\,212.8 and 243.2 MeV for the $^{48}$Ca+$^{254}$Fm and $^{54}$Cr+$^{248}$Cm
systems, respectively. Notice that the actual threshold energies for fusion are somewhat smaller
than these values due to the dynamical modifications of the barriers \cite{Dasgupta98,BT98,HT12,
Back14,Montagnoli17}. The figure also shows the density distributions for each system at $R=R_{\rm min}$,
11\,fm, 12\,fm, and 13\,fm, where $R_{\rm min}$ is the distance of the closest approach.
The figure indicates that the energy loss in the approaching phase is indeed larger for the
$^{54}$Cr+$^{248}$Cm system than for the $^{48}$Ca+$^{254}$Fm system.
However, the distance of the closest approach, $R_{\rm min}$, does not differ much, that is,
$R_{\rm min}$\,=\,10.33 and 10.40 fm for the $^{48}$Ca+$^{254}$Fm and $^{54}$Cr+$^{248}$Cm
systems, respectively, for the case of $E=1.2V_b$ (see the arrows in Fig. \ref{fig1}).
At $E=V_b$, the distance of the closest approach is $R_{\rm min}$\,=\,12.93 and 13.09 fm
for the former and the latter systems, respectively. This implies that the magicity of the $^{48}$Ca
nucleus plays a minor role in determining the distance of the closest approach, even though 
it significantly affects the dynamics before the touching. 

Here, we mention that the discontinuity in the energy loss at small relative distances shown in Fig.~\ref{fig1}
is due to a tiny jump of a neck position, which causes a discontinuity in $V(R(t))$ and $\gamma(R(t))$
through a numerical time derivative. Even though we may remedy it by improving the neck detection 
algorithm, the conclusion of the present paper will be maintained, as it does not change much the
value of $R_{\rm min}$.

Let us next evaluate the diffusion probability, $P_{\rm CN}$, and the survival probability, $W_{\rm sur}$,
in Eq.~(\ref{er}) using the distances of the closest approach evaluated with the TDHF calculations. 
To this end, we employ the fusion-by-diffusion model \cite{fbd03,Hagino18,fbd11,fbd12}. In this model,
the diffusion process is described as a diffusion over a simple one-dimensional parabolic potential from
an injection point \cite{Abe00}, while the decay of the compound nucleus is described with a simplified 
statistical model in which only the competitions between fission and neutron evaporations are taken into
account. In the overdamped regime assumed in the fusion-by-diffusion model, the diffusion probability
depends only on the temperature and the height of the inner barrier \cite{fbd03,Abe00}. Therefore,
after a mass formula, a level density parameter, and a parametrization of the inner barrier are specified,
there remains only a single adjustable parameter, i.e., the injection point parameter, which defines the
effective height of the inner barrier. We estimate the injection point parameter as, 
\begin{equation}
s_{\rm inj}=R_{\rm min}-R_{\rm P}-R_{\rm T},
\end{equation}
where $R_{\rm min}$ is the distance of the closest approach obtained from the TDHF calculations,
while $R_{\rm P}$ and $R_{\rm T}$ are the radii of the projectile and the target nuclei, respectively.
In this paper, we use 1.15 fm for the radius parameter. The explicit form of the inner barrier is given in
Ref.~\cite{fbd11}, for which we use the default parameter set. Notice that the deformation effect is taken
into account in the extended fusion-by-diffusion model through the orientation angle dependence of
$R_{\rm min}$, while the inner potential remains independent of the orientation angle \cite{Hagino18}.
We also assume that the kinetic energy is completely dissipated to the internal energy at the injection point.
See Refs.~\cite{fbd03,Hagino18,fbd11,fbd12} for other details of the fusion-by-diffusion model.

\begin{table}[bt]
\caption{
The diffusion probability, $P_{\rm CN}$, the survival probability, $W_{\rm sur}$, and their product,
$P_{\rm CN}W_{\rm sur}$, estimated with the fusion-by-diffusion model for several hot fusion reactions
leading to the element 120. These quantities are evaluated for the $s$-wave scattering at $E=V_b$ for
the side collision for each reaction. The excitation energy, $E^*$, of the compound nucleus (CN) as well as
the distance of the closest approach estimated with the TDHF calculations are also shown.
}
\begin{center}
\begin{tabular}{c|cccccc}
\hline
\hline
System & CN & $E^*$ & $R_{\rm min}$ & $P_{\rm CN}$ & $W_{\rm sur}$ & $P_{\rm CN}\,W_{\rm sur}$ \\
& & (MeV) & (fm) & ($\times 10^4$) & ($\times 10^9$) & ($\times 10^{13}$) \\
\hline
$^{48}$Ca+$^{254}$Fm & $^{302}$120 & 29.0 & 12.93 & 1.72 & 176 & 302 \\
$^{54}$Cr+$^{248}$Cm & $^{302}$120 & 33.2 & 13.09 & 1.89 & 1.31 & 2.47 \\
$^{51}$V+$^{249}$Bk & $^{300}$120 & 37.0 & 12.94 & 3.95 & 0.117 & 0.461 \\
$^{48}$Ca+$^{257}$Fm & $^{305}$120 & 30.5 & 12.94 & 2.49 & 0.729 & 1.82 \\
\hline
\hline
\end{tabular}
\end{center}\label{table}
\end{table}

Table~\ref{table} summarizes the diffusion and survival probabilities evaluated  at $E=V_b$.
Here, the survival probabilities represent the total survival probabilities, which are a sum of probabilities
for all neutron evaporation channels. Following Refs.~\cite{Hagino18,fbd12}, we use the mass formula
and the fission barrier heights given in Ref.~\cite{Kowal12}. This mass formula predicts the fusion
$Q$ values of $Q_{\rm fus}$\,=\,$-183.8$ and $-210.10$ MeV for the $^{48}$Ca+$^{254}$Fm
and the $^{54}$Cr+$^{248}$Cm systems, respectively, and thus the excitation energy of the compound
nucleus formed at $E=V_b$ in the former reaction is smaller than that formed in the latter reaction.
Because the distances of the closest approach are similar to each other, the diffusion probability, $P_{\rm CN}$,
for the $^{54}$Cr+$^{248}$Cm reaction is slightly larger than that for the $^{48}$Ca+$^{254}$Fm reaction,
reflecting the higher excitation energy. On the other hand, the survival probability, $W_{\rm sur}$, is much
more sensitive to the excitation energy, and that for the $^{48}$Ca+$^{254}$Fm reaction is larger than
that for the $^{54}$Cr+$^{248}$Cm reaction by about two orders of magnitude. The products of
the diffusion and survival probabilities are also different by a similar amount. This clearly indicates that
the main effect of the magicity of the $^{48}$Ca nucleus is due to the low excitation energies of the compound
nucleus, whereas the dynamics of the entrance channel plays a much less important role. This is in contrast
to the cases with heavy magic nuclei, for which the magicity plays an important role in the entrance channel
dynamics \cite{Satou06,Moller97}. 

\begin{figure} [tb]
\includegraphics[width=8.6cm]{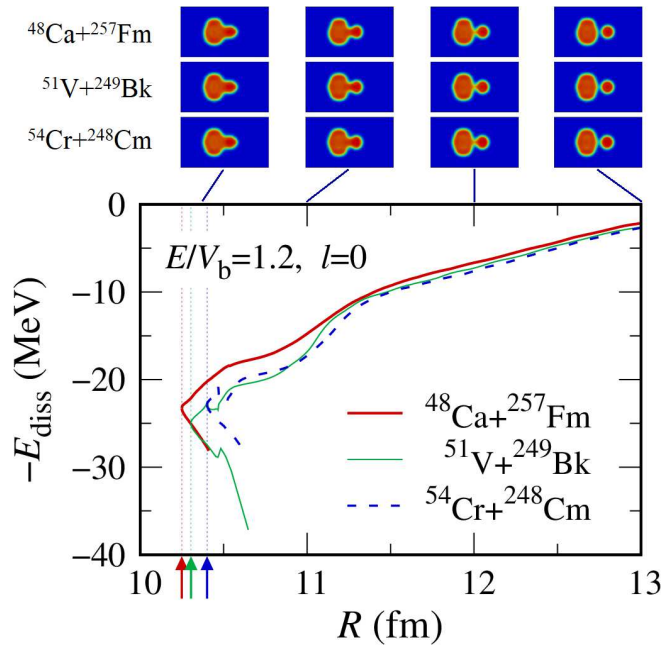}
\caption{
Same as Fig.~\ref{fig1}, but for the $^{48}$Cr+$^{257}$Fm (the thick solid line),
the $^{51}$V+$^{249}$Bk (the thin solid line), and the $^{54}$Cr+$^{248}$Fm
(the dashed line) systems.
}
\label{fig2}
\end{figure}

We next discuss the projectile-target combinations which are experimentally accessible,
namely, the $^{54}$Cr+$^{248}$Cm\,$\to ^{302}$120 and the $^{51}$V+$^{249}$Bk\,$\to ^{300}$120
systems. For comparison, we also consider the $^{48}$Ca-induced reaction with $^{257}$Fm,
which is the longest lived Fm isotope (with the half-life of 100.5 days), that is, $^{48}$Ca+$^{257}$Fm\,$
\to ^{305}$120. The energy losses for the relative motion at $E=1.2V_b$ evaluated with the TDHF
calculations are shown in Fig.~\ref{fig2} for these three systems. For the $^{51}$V nucleus, we apply 
the filling approximation for protons in the $f_{7/2}$ orbitals in order to obtain a spherical ground state.
The barrier heights are estimated to be $V_b$\,=\,237.0 and 212.4 MeV for the $^{51}$V+$^{249}$Bk
and the $^{48}$Ca+$^{257}$Fm systems, respectively. The energy losses for these systems are
qualitatively the same as those shown in Fig.~\ref{fig1}. That is, even though the energy loss for the
$^{48}$Ca-induced reaction is somewhat lower than the energy losses for the other two reactions,
the distances of the closest approach are similar to each other among the three systems. The distance of
the closest approach is estimated to be 10.25, 10.30, and 10.40 fm for the $^{48}$Ca+$^{257}$Fm,
the $^{51}$V+$^{249}$Bk, and the $^{54}$Cr+$^{248}$Cm systems, respectively. The distances
of the closest approach evaluated at $E=V_b$ are also summarized in Table~\ref{table}.

The diffusion and the survival probabilities for $E=V_b$ for each system are shown in Table~\ref{table}.
The small total probability, $P_{\rm CN}W_{\rm sur}$, for the $^{51}$V-induced reaction
is caused by the high excitation energy for this system. In addition, an interesting observation
is that the survival probability for the $^{48}$Ca+$^{257}$Fm reaction is smaller than that for the 
$^{54}$Cr+$^{248}$Cm reaction by a factor of about 1.8 despite the fact that the excitation energy
is smaller by about 3 MeV. This is in marked contrast to the comparison to the $^{48}$Ca+$^{254}$Fm
reaction, for which the low excitation energy (by about 4 MeV) enhances the survival probability
by about two orders of magnitude. This difference originates mainly from the mass number dependence
of the fission barrier height. The fission barrier heights for the element $Z=120$ evaluated in Ref.~\cite{Kowal12}
are 5.04, 4.66, and 3.54 MeV for $A$\,=\,300 ($^{51}$V+$^{249}$Bk), 302 ($^{54}$Cr+$^{248}$Cm
and $^{48}$Ca+$^{254}$Fm), and 305 ($^{48}$Ca+$^{257}$Fm), respectively (Ref.~\cite{Kowal12}
provides the results for even-even nuclei only, and for the $^{305}$120 nucleus we have taken an
average of the fission barrier heights for $A$\,=\,304 and 306). That is, the fission barrier height for
the compound nucleus formed in the $^{48}$Ca+$^{257}$Fm reaction is low, leading to the low
survival probability. This again implies that the magicity of the $^{48}$Ca nucleus plays a minor role
in the entrance channel dynamics in reactions forming superheavy elements.

In summary, we have developed a novel approach for fusion reactions for superheavy elements.
This combines good aspects of the microscopic time-dependent Hartree-Fock (TDHF) method and
a phenomenological Langevin approach for the diffusion process over the inner barrier: we have
used the TDHF approach for the entrance channel dynamics in order to estimate the distance of
the closest approach without an empirical parameter, which provides the initial condition for the Langevin
approach. We have applied this new approach to several systems which lead to synthesis of the element 120,
i.e., $^{48}$Ca+$^{254,257}$Fm, $^{51}$V+$^{249}$Bk, and $^{54}$Cr+$^{248}$Cm
reactions. We have shown that the distances of the closest approach are similar to one another as
long as the incident energy relative to the Coulomb barrier height for each system is kept to be the
same. The magicity of the $^{48}$Ca nucleus thus influences mainly the evaporation process
through the excitation energies of the compound nuclei. We have found that the formation probability
of evaporation residues for $^{48}$Ca+$^{254}$Fm\,$\to ^{302}$120 is larger than that for 
$^{54}$Cr+$^{248}$Cm\,$\to ^{302}$120 by about two orders of magnitude. On the other hand,
the probability for the latter reaction is larger than that for $^{51}$V+$^{249}$Bk\,$\to^{300}$120
by a factor of about 5, reflecting the difference in the excitation energies of the compound nuclei.
Despite the magicity of the $^{48}$Ca projectile, the probability for the $^{48}$Ca+$^{257}$Fm\,$\to^{305}$120
reaction has been found to be slightly smaller than that for the $^{54}$Cr+$^{248}$Cm\,$\to^{302}$120
reaction, due to a low fission barrier height of the $^{305}$120 nucleus.

In this paper, for simplicity, we have considered only the $s$-wave scattering for the side collision geometry.
In order to compute evaporation residue cross sections, one would need to add contributions of other partial
waves and also to take an average over the orientation angles of the deformed target nuclei. We would,
however, not expect that our conclusions in this paper will be altered qualitatively. Also, we have neglected
the pairing correlations in the TDHF calculations. The pairing correlations may affect reaction dynamics,
especially for the cases with open-shell projectiles, e.g., $^{50}$Ti, $^{51}$V, and $^{54}$Cr, in a way as was
discussed recently in Refs.~\cite{HS16,MSW17}. This is completely an open issue, which should be addressed
separately in the future. Another simplification which we have taken in this paper is that we have used a simple
fusion-by-diffusion model for the diffusion process. This can be improved by using more sophisticated Langevin
calculations. We are currently working on this, and we will report it in a separate publication. Another interesting
application of the present approach is to the $^{208}$Pb based cold fusion reactions, for which the magicity
of the $^{208}$Pb nucleus plays an important role in the entrance channel dynamics. We leave it for a future work.

\medskip

We thank K. Washiyama and Y. Aritomo for useful discussions.
This work used computational resources of the HPCI system (Oakforest PACS)
provided by the Joint Center for Advanced High Performance Computing (JCAHPC)
through the HPCI System Research Projects (Project No. hp180080).
This work also used (in part) computational resources of the COMA (PACS-IX)
System provided by Multidisciplinary Cooperative Research Program in Center
for Computational Sciences, University of Tsukuba.

\end{document}